\documentclass[pre,twocolumn,a4paper,superscriptaddress,floatfix]{revtex4}
\usepackage{graphicx,amssymb}
\usepackage[utf8]{inputenc}
\usepackage[T1]{fontenc}
\usepackage{colortbl}
\usepackage{xcolor}

\begin{document}

\newcommand{\be}{\begin{equation}}
\newcommand{\ee}{\end{equation}}
\newcommand{\bq}{\begin{eqnarray}}
\newcommand{\eq}{\end{eqnarray}}
\newcommand{\bsq}{\begin{subequations}}
\newcommand{\esq}{\end{subequations}}
\newcommand{\bc}{\begin{center}}
\newcommand{\ec}{\end{center}}
\newcommand {\R}{{\mathcal R}}
\newcommand{\al}{\alpha}
\newcommand\lsim{\mathrel{\rlap{\lower4pt\hbox{\hskip1pt$\sim$}}
    \raise1pt\hbox{$<$}}}
\newcommand\gsim{\mathrel{\rlap{\lower4pt\hbox{\hskip1pt$\sim$}}
    \raise1pt\hbox{$>$}}}

\title{Mobility-limiting antipredator response in the rock-paper-scissors model}

\author{J. Menezes}  
\email[Electronic address: ]{jmenezes@ect.ufrn.br} 
\affiliation{Escola de Ci\^encias e Tecnologia, Universidade Federal do Rio Grande do Norte\\
Caixa Postal 1524, 59072-970, Natal, RN, Brazil}
\affiliation{Institute for Biodiversity and Ecosystem Dynamics, University of Amsterdam, Science Park 904, 1098 XH Amsterdam, The Netherlands}

\author{B. Moura}  
\email[Electronic address: ]{bianpmoura@gmail.com} 
\affiliation{Departamento de Engenharia Biomédica, Universidade Federal do Rio Grande do Norte\\
Av. Senador Salgado Filho, 300, 59078-970, Natal, RN, Brasil}
\affiliation{Edmond and Lily Safra International Neuroscience Institute, Santos Dumont Institute\\
Av Santos Dumont, 1560, 59280-000, Macaiba, RN, Brazil}

\date{\today}
\begin{abstract}
Antipredator behavior is present in many biological systems where individuals collectively react to an imminent attack.
The antipredator response may influence spatial pattern formation and ecosystem stability but requires an organism's cost to contribute to the collective effort. We investigate a nonhierarchical tritrophic system, whose predator-prey interactions are described by the rock-paper-scissors game rules. In our spatial stochastic simulations, the radius of antipredator response defines the maximum prey group size that disturbs the predator's action, determining the individual cost to participate in antipredator strategies. We consider that
each organism contributes equally to the collective effort, having its mobility limited by the proportion of energy devoted to the antipredator reaction. Our outcomes show that the antipredator response leads to spiral patterns, with the segregation of organisms of the same species occupying departed spatial domains. We found that a less localized antipredator response increases the average size of the single-species patches, improving the protection of individuals against predation. Finally, our findings show that although the increase of the predation risk for a more localized antipredator response, the high mobility constraining benefits species coexistence.
Our results may help ecologists understand the mechanisms leading to the stability of biological systems where locality is crucial to behavioral interactions among species.
\end{abstract}
\maketitle

\section{Introduction}

There is plenty of evidence that spatial segregation of species is fundamental to the formation and stability of ecosystems \cite{ecology,Nature-bio, BUCHHOLZ2007401}. For example, experiments with bacteria \textit{Escherichia coli} revealed the role of space in the preservation of biodiversity \cite{Coli}. The authors demonstrated that
the cyclic dominance among three bacteria strains could be described by the spatial rock-paper-scissors game rules \cite{Allelopathy}. However, they observed that the cyclic selection interactions were not sufficient to guarantee coexistence unless individuals interact locally. Their studies revealed that the spatial interactions result in departed spatial domains occupied by individuals of the same species; similar spatial patterns appear in groups of lizards, and coral reefs \cite{lizards,Extra1}.

It is well known that behavioral strategies play a vital role in evolutionary biology \cite{BUCHHOLZ2007401}. For example, movement strategies may be decisive to the success of individuals to guarantee
natural resources or refuges against enemies \cite{Moura,Motivation1,butterfly,refuge1,refuge2}. 
Another common animal behavior is the resistance against predation \cite{DefenseAnimals}. It has been reported that vertebrates and invertebrates perform an antipredator tactic called Thanatonis, i.e., death feigning \cite{Thanatonis,AntiFish1}. Prey mites \textit{Tetranychus urticae} emit an odor when exposed to the predatory mite \textit{\mbox{Phytoseiulus} \mbox{persimilis}} to reduce the oviposition, and the consequent predator population growth \cite{ContraAtacck2}. Several other examples of antipredator behavior of mites, like the variation of the nest size and web density have been studied \cite{NinhosdeMites2,MitesWeb1,MitesWeb2}. It has been reported that antipredator behavior leads individuals to join efforts to respond to predation threats \cite{Grouping1,Grouping2,LizardB1}. The herd behavior allows individuals to be less vigilant for imminent attacks because grouping increases the probability of predator detection, which may stabilize the predator-prey system at a population level \cite{herd1,detection,NinhosdeMites1}. Furthermore, it has been reported that  as the prey group size increases, more eyes oversee the environment, increasing the collective response to an imminent onslaught from any predator that approaches the group \cite{vigilance3,vigilance2}.
Although there are benefits of the collective defense strategies, there is a cost associated with the antipredator behavior that brings consequences, for example, on the individual mobility \cite{Cost2,Cost3,Cost0,Cost1}. The cost depends on the individual
effort expended by a single prey to contribute to the collective
antipredator activity, decreasing as the number of collaborators grows \cite{vigilance4,vigilance5}.

Stochastic simulations of the rock-paper-scissors game have been a tool to comprehend how spatial patterns influence biodiversity in cyclic models \cite{Szolnoki_2020, Szolnoki-JRSI-11-0735}. There are two implementation versions: with or without a conservation law for the total number of individuals. Namely, Lotka-Volterra \cite{weakest,doi:10.1021/ja01453a010,Volterra,PhysRevE.78.031906,0295-5075-121-4-48003,basins,PhysRevE.82.066211} and May-Leonard realizations \cite{Reichenbach-N-448-1046,Avelino-PRE-86-031119,Avelino-PRE-86-036112,uneven}, respectively. In the Lotka-Volterra implementation - where the interactions are predation and random mobility - spiral patterns are formed when prey respond to the predator's action. According to the outcomes presented in Ref. \cite{Anti1}, the spiral waves appear because the predation mostly happens on the borders of predator-dominated spatial domains. Moreover, the strength of the antipredator reaction controls the characteristic length of the spatial patterns and the coexistence probability.

\begin{figure}
\centering
\includegraphics[width=45mm]{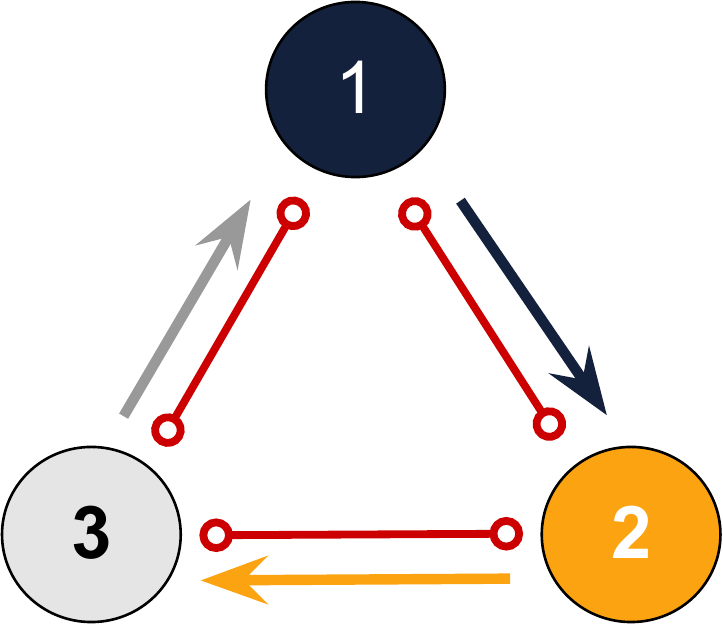}
\caption{Illustration of interaction rules in our cyclic tritrophic system. Red solid lines indicate that organisms of every species compete equally for space. Black, orange, and gray arrows illustrate the cyclic predator-prey interactions.}
	\label{fig1}
\end{figure}

In this paper, we investigate a mobility-limiting local antipredator response in nonhierarchical tritrophic systems described by the rock-paper-scissors game rules. We introduce a cost for an organism to perform antipredator behavior that limits its mobility probability. The cost depends on the radius of the antipredator response that determines the maximum prey group size that can influence the predator's action. We study how the locality of the antipredator response influences predation risk, and consequently, the size of the spatial domains. Our main goal is to comprehend what changes if the local antipredatory response is limited to individuals located at different distances from the predator. In our model, 
i) in the imminence of a predator's attack, individuals of every species have the same strength to resist predation; 
ii) whenever an individual is in danger, it counts on its conspecifics to react to the predation threat; 
iii) the radius of the antipredator response limits the maximum prey group size disturbing the predator action;
iv) antipredator action involves permanent vigilance to scan the environment aiming to detect predator presence;
v) the cost for an organism to perform antipredator behavior
is the same for every species and depends on the maximum prey group that can join the collective resistance;
vi) the cost for antipredator defense tactics constrains the individuals' mobility probability.
We aim to discover how the radius of the antipredator response impacts the population dynamics and jeopardizes biodiversity.

The outline of this paper is as follows. In Sec. II, we introduce the stochastic rules of the tritrophic system with antipredator behavior. In Sec. III, we focus on the changes of the spatial patterns for several radii of the antipredator response. In Sec. IV, we investigate the dynamics of the spatial densities, while the impact of the antipredator reaction on an individual's predation risk appears in Sec. V.
The analysis of the autocorrelation function is realized in Sec. VI, whereas the coexistence probability in terms of the individual's mobility is addressed in Sec. VII. Finally, our comments and conclusions appear in Sec. VIII.

\label{Introduction}

\begin{figure}
\centering
\includegraphics[width=45mm]{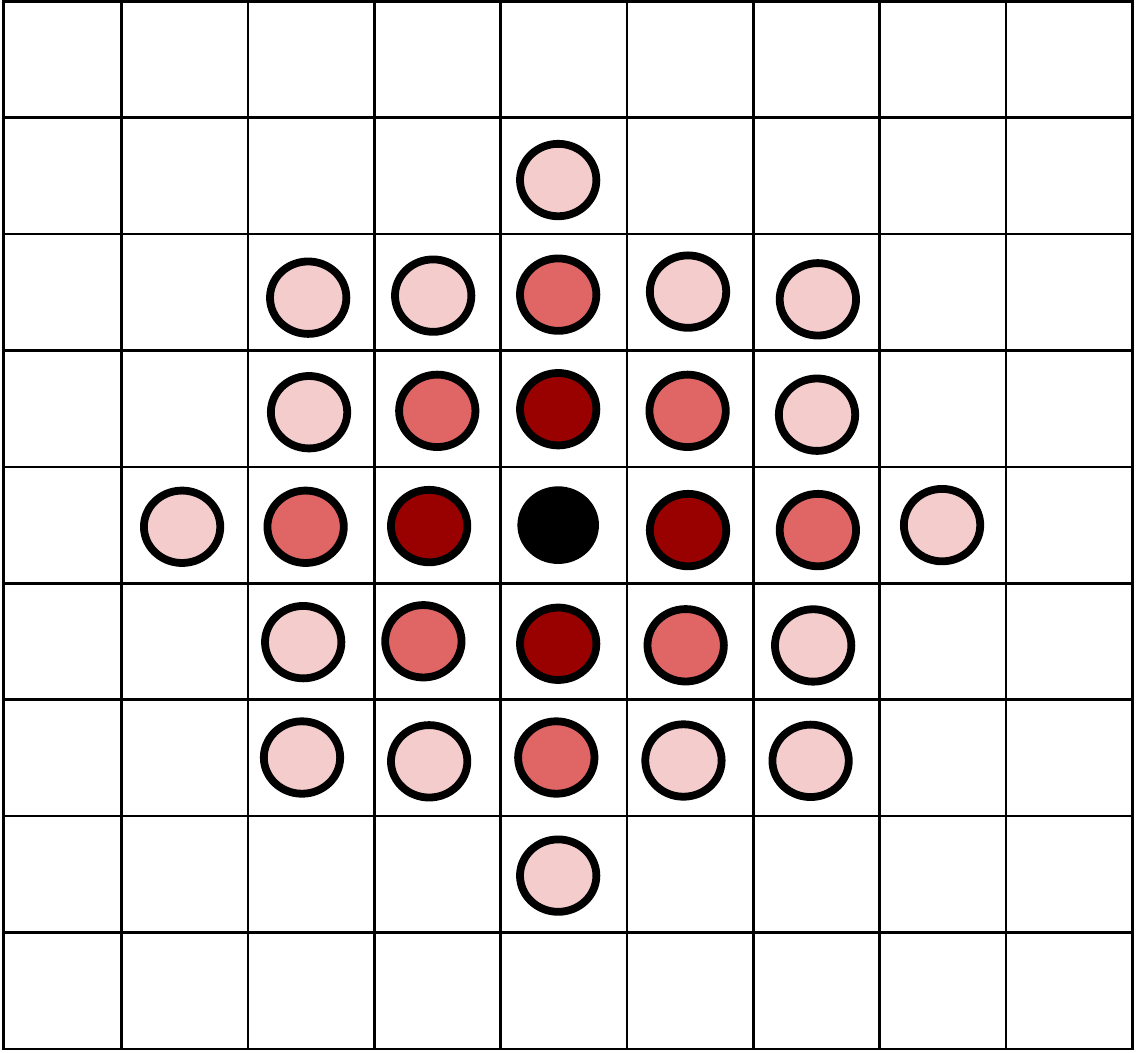}
\caption{Illustration of the radius of the antipredator response in a predator's neighborhood. Ruby indicates the grid sites for $R=1$; ruby and red show the positions for $R=2$; ruby, red, and pink dots form the set of grid sites for $R=3$. The black dot represents the predator.}
	\label{fig2}
\end{figure}

\begin{figure*}[t]
	\centering
	\includegraphics*[width=3.8cm]{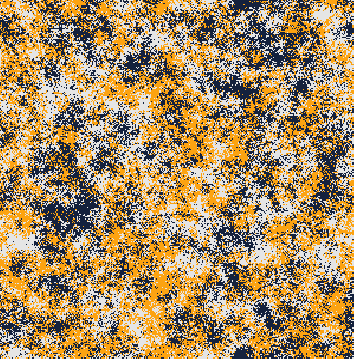}
	\includegraphics*[width=3.8cm]{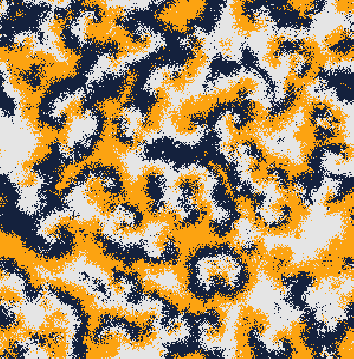}
	\includegraphics*[width=3.8cm]{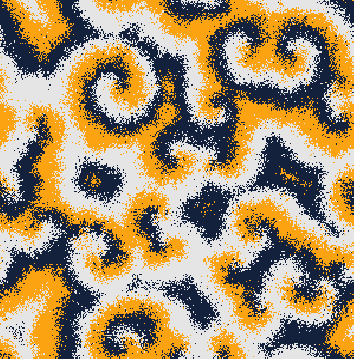}
	\includegraphics*[width=3.8cm]{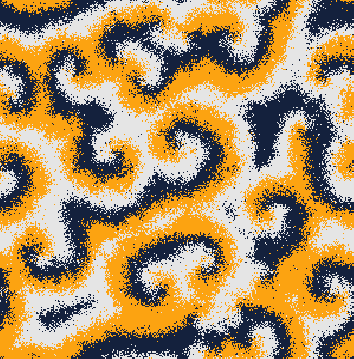}
\caption{Snapshots of simulations of the rock-paper-scissors game illustrated in Fig. \ref{fig1} running in square lattices with $300^2$ grid points. Each dot shows an individual according to the color scheme in Fig. \ref{fig1}. From left to right, the panels show the results for the standard model, $R=1$, $R=3$, and $R=5$, respectively. The simulations started from the same random initial conditions.}
 \label{fig3}
\end{figure*}
\section{The Model}

We investigate a tritrophic system where species dominate each other according to the popular rock-paper-scissors game rules. Species are labeled by $i$ with $i= 1,...,3$, with the cyclic identification $i=i\,+\,3\,\alpha$ where $\alpha$ is an integer. Accordingly, organisms of species $i$ prey upon individuals of species $i+1$. In our model, individuals of every species perform antipredator behavior: a prey group surrounding the predator opposes predation, causing a decrease in the predation probability that depends on the group size and the strength of the antipredator response.
The cost of the antipredator behavior depends on the maximal size of the prey group that may join the effort in the reaction; this is 
modeled by reducing the individuals' probability mobility, and it depends on the maximal size of the prey group that may join the effort in the reaction. 

Our stochastic simulations are performed in square lattices with periodic boundary conditions. We assume a conservation law for the total number of individuals, following the Lotka-Volterra numerical implementation of the rock-paper-scissors game \cite{doi:10.1021/ja01453a010,Volterra}. Each grid point contains one individual; thus, the total number of individuals is always equal to $\mathcal{N}$, the total number of grid points. 

The possible interactions are:
\begin{itemize}
\item 
Predation: $ i\ j \to i\ i\,$, with $ j = i+1$. When one predation interaction occurs, a organism of species $i$ (the predator) replaces the grid point filled by the 
by the individual of species $i+1$ (the prey).
\item
Mobility: $ i\ \odot \to \odot\ i\,$, where $\odot$ means an individual of any species. When moving, an individual of species $i$ switches positions with another organism of any species.
\end{itemize}
Figure $1$ illustrates our stochastic model's predation and mobility rules. The arrows indicate a cyclic trophic dominance among the species; the solid lines show that species equally compete for space.


To explore the local aspects of antipredator behavior, we define the 
the radius of the antipredator response $R$ as the maximum distance from the predator at which prey can interfere with the predator's action. Figure 2 illustrates the numerical implementation of the radius of the antipredator reaction in a predator's neighborhood. For $R=1$, a predator, located at the black dot, feels the opposition only of prey in the ruby grid sites; predation is impacted by the reaction of prey in the ruby and red positions for $R=2$; in the case of $R=3$, prey in ruby, red, and pink dots can disturb predation. The prey effort may be devoted to surveillance or participation in any collective defense strategy against the predator. This behavior has an intrinsic individual cost $c$ that is demanded of each organism, defined as the fraction of the maximum antipredator response suffered by a predator \cite{vigilance2,vigilance3,vigilance4}.
The descending colors in Fig. 2 indicate the decreasing individual effort to the collective antipredator resistance against predation. Specifically, for $R=1$, $R=2$, and $R=3$, each individual contributes with $c\,=\,1/4$, $c\,=\,1/12$, and $c\,=\,1/28$ of the maximal antipredator response, respectively.

Therefore, for a given predator of species $i$, the effective predation probability is a function of the fraction of individuals of species $i+1$ within a disk of radius $R$, centered at the predator is 
\be
p\,=\,p_0\,e^{-\kappa\,\frac{\mathcal{G}}{\mathcal{G}_{max}}}
\label{eq1}
\ee
with $p_0$ being the predation probability in the standard model, without antipredator behavior.
In Eq.~\ref{eq1}, $\mathcal{G}_{max}$ is the maximum group size (the number of individuals that fit within a disk of radius $R$); $\mathcal{G}$ is the actual group size; $\kappa$ is the antipredator strength factor, a real parameter defined as $\kappa\geq0$, where $\kappa=0$ represents the standard model, that is, $p=p_0$. 

The individual cost of a lonely prey to contribute to collective antipredator response is $c=1/\mathcal{G}_{max}$.
In this scenario, for $\kappa\,>\,0$, a lonely prey's opposition reduces the predation probability to $p=p_0\,e^{-c\,\kappa}$ while $p$ is minimal when $\mathcal{G\,}=\,\mathcal{G}_{max}$, i.e., $p=p_0\,e^{-\kappa}$. We investigate the local effects of the antipredator response for $1\,<\,R\,<\,5$, where $R$ is measured in units of the lattice spacing.
Furthermore, as we are interested in understanding the effects of the locality of the antipredator response, we assume a fixed $\kappa=2.0$ so that our results are independent of the strength of the antipredator reaction. However, we have verified that the main conclusions presented in this paper hold for other values of $\kappa$.

Our simulations begin with random initial conditions, where each grid point is given an organism of an arbitrary species. Initially, the total numbers of individuals of every species are the same: $I_i=\mathcal{N}/3$, for $i=1,2,3$. The interactions were implemented with the Moore neighborhood, i.e., individuals may interact with one of their eight immediate neighbors. The simulation algorithm follows three steps: i.) randomly selecting an individual to be the active one; ii.) drawing one of its eight neighbor sites to be the passive individual; iii.) randomly choosing an interaction to be executed by the active individual. One timestep is counted if the active and passive individuals (steps i and ii) allow the selected interaction interaction (step iii) to be executed. Otherwise, the three steps are repeated. Our time unit is called generation, which is the necessary time to $\mathcal{N}$ interactions to occur. 

In the absence of the antipredator behavior, predation and mobility probabilities are denoted by $m_0$ and $p_0$, respectively, with $m_0\,+p_0\,=\,1$. Nonetheless, if organisms perform antipredator response, two changes are implemented. First, because of the cost of the antipredator strategy, mobility probability is limited due to the individual cost of cooperating with the group: $m_0 = c + m$, where $m$ is the effective mobility probability. Second, the effect of the collective antipredator response reduces the chances of a predation interaction being implemented according to Eq. \ref{eq1}: $p < p_0$.

\section{Spatial Patterns}

\begin{figure}[t]
\centering
\includegraphics[width=50mm]{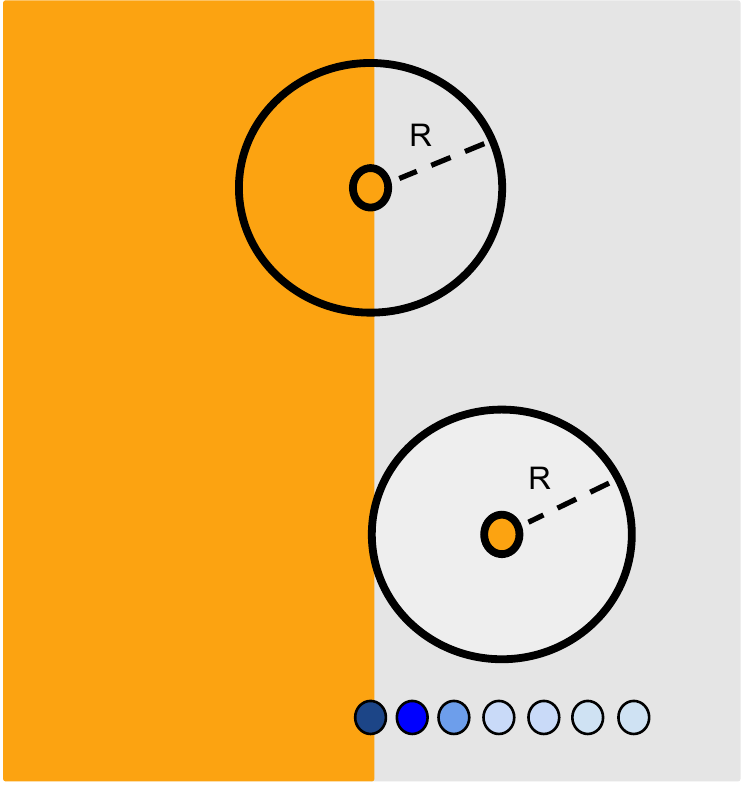}
\caption{Illustration of prey group size around a predator.  Gray and orange dots indicate predators and prey, respectively. The radius of the antipredator response $R$ shows the maximum distance a prey can interfere with the predator, whereas the blue descending circles indicate the decrease in predation probability as the predator moves away from the domain interface.}
	\label{fig4}
\end{figure}

To observe the spatial patterns, we first run a single realization for
different values of $R$ to be compared with the standard model (without antipredator behavior). The simulations run in square lattices with $300^2$ sites for a timespan of $3000$ generations, assuming $p\,=\,m\,=\,1/2$. From left to right, the snapshots in Fig. 3 show the spatial patterns at the end of the simulations for the standard model, $R=1$, $R=3$, and $R=5$, respectively. The colors follow the scheme in Fig.~\ref{fig1}: black, orange, and gray dots depict individuals of species $1$, $2$, and $3$, respectively.

In the absence of antipredator opposition, predators consume prey everywhere. This provokes a continuous change in the local species segregation, as depicted in the first panel of Fig. 3 that shows irregular groups of individuals of the same species. If organisms perform antipredator response, each predator faces its local reality:
the larger the prey group size in the predator's vicinity, the more difficult consuming the prey. Therefore, predators positioned close to conspecifics have more chances of feeding, which causes the arising of spirals: individuals of the same species congregate in spatial domains forming the spiral arms \cite{Anti1}.

To understand the influence of the radius of the antipredator response on spiral patterns, let us suppose a spatial domain of species $2$ (predator) invading a region inhabited by individuals of species $3$ (prey), as illustrated in Fig. 4. In this hypothetical case, a predator on the border of the spatial domains faces the opposition of a prey group with size $\mathcal{G}\,=\mathcal{G}_{max}/2$. In contrast, another predator distant from $R$ grid sites to the border copes with the resistance of a prey team with size $\mathcal{G}\,=\mathcal{G}_{max}$. This means that as a predator moves away from the orange area, its effective predation probability decreases, as illustrated by the blue descending circles on the bottom of Fig. 4. Namely, $p_{eff}$ varies from $p/e$ on the border to $ p/e^2$ on distances equal or larger than $R$. 

When it comes to the spiral patterns in Fig. 3, a predator deals with an increasing antipredator response when moving away from the border between two spiral arms. However, if $R$ increases, the predator can go further, struggling with less prey resistance. This means that, for larger
$R$, the antipredator response is less localized, making it less difficult for a predator away from the boundaries of predator-dominated domains to consume prey. The consequence is the increase of the average spatial domain size observed in the third and fourth panels of Fig. 3 ($R=3$ and $R=5$, respectively) compared to the second one ($R=1$). 
\section{Dynamics of Species Densities}

We calculate the spatial species densities $\rho$, i.e., the fraction of the grid occupied by individuals of the species $i$ in the single realizations shown in Fig. 3. For this, we focus only on the spatial density of species $1$, which is a function of time $t$, i.e., $\rho(t) = I_1(t)/\mathcal{N}$, because of the tritrophic chain's symmetry of the rock-paper-scissors model, the average spatial densities are the same irrespective of the species. 

Figure $5$ shows the dynamics of the species densities
in the simulations presented in Fig 3. The gray line shows the dynamics of $\rho$ for the standard model while the yellow, blue, and red lines represent the results for $R=1$, $R=3$, and $R=5$, respectively. Our findings reflect the cyclic territorial dominance of species $i$ ($i=1,2,3$), characteristic of the rock-paper-scissors models \cite{doi:10.1021/ja01453a010,Volterra}. The growth in the average spatial domain size is responsible for increasing the amplitude and frequency of the species densities for larger $R$.

\begin{figure}
	\centering
	\includegraphics*[width=8.5cm]{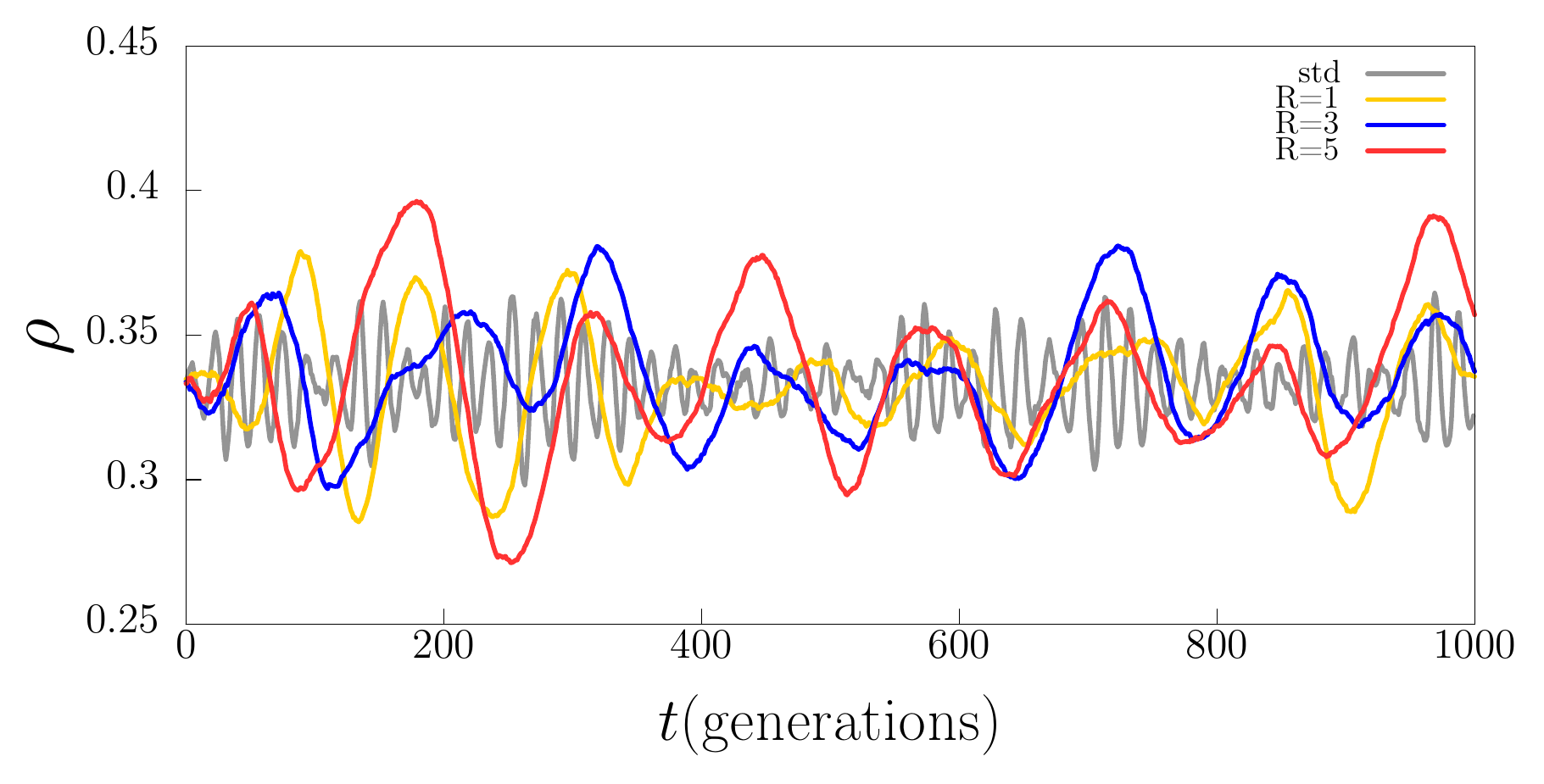}
\caption{Temporal changes of spatial species densities $\rho$ in the simulations presented in Fig.~\ref{fig2}. The gray, yellow, blue, and red lines represent the results for antipredator strength factor $\kappa=0$ (standard model), $R=1$, $R=3$, and $R=5$, respectively. }
 \label{fig5}
\end{figure}

\section{Predation Risk}
\begin{figure}[t]
	\centering
	\includegraphics*[width=8.6cm]{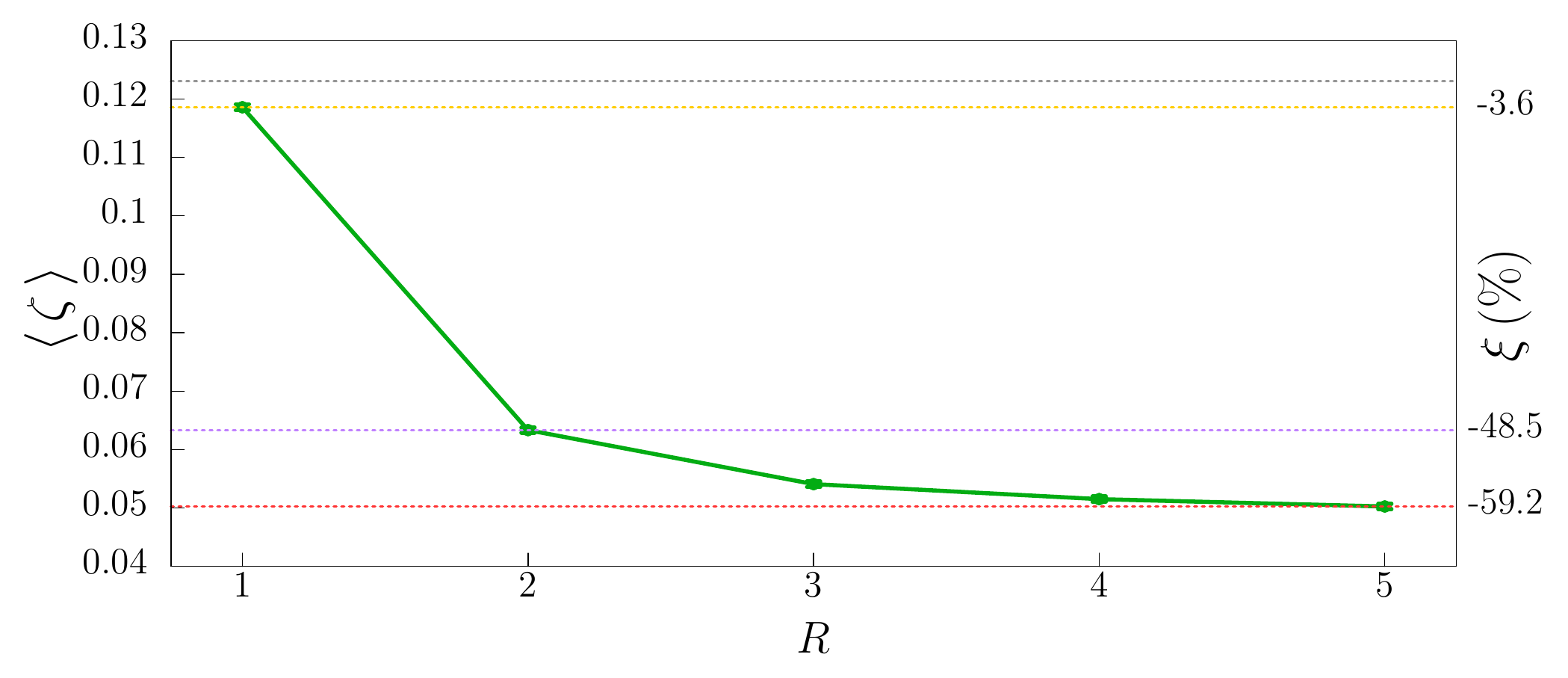}
\caption{Mean predation risk $\langle\,\zeta\,\rangle$ in terms of the radius of the antipredator response $R$. The results were averaged from a set of $100$ simulations of squares lattices with $300^2$ points. The right axis shows the relative change in the predation risk in comparison with the standard model (horizontal gray dashed line). The horizontal yellow, purple, and red dashed lines depict $\,\xi\,(\%)$ for $R=1$, $R=2$, and $R=5$, respectively.
}
 \label{fig6}
\end{figure}
We aim to understand how the risk of an organism being consumed depends on the radius of the antipredator response. Having assumed the same predation probability $p_0$ for every species, we focus on computing the predation risk for species $1$. For this purpose, we first counted the total number of individuals of species $1$ at the beginning of each generation. Subsequently, we computed how many individuals of species $1$ are preyed on during the generation. We define the predation risk, $\zeta$, as the ratio between the number of consumed individuals and the initial amount. To avoid the noise inherent in the pattern formation stage, we calculated the predation risk considering only the second half of the simulation. Also, we averaged the results every $30$ generations. 

We performed $100$ realizations with different random initial conditions for each value of $R$. The mean value of the predation risk, $\langle\,\zeta\,\rangle$ is depicted in Fig. $6$ for $1\leq\,R\,\leq\,5$, where the error bars show the standard deviation. Figure $6$ also shows the percentage relative predation risk $\xi =(\langle\,\zeta\,\rangle - \langle\,\zeta_0\,\rangle)/\langle\,\zeta_0\,\rangle$, where $\zeta_0$ is the predation risk in the absence of the antipredator behavior: $\zeta_0= 0.123$ (Ref.~\cite{Anti1}).
The outcomes revealed that the predation risk reduction is $3.6\%$ (yellow dashed line) for $R=1$ in comparison with the standard model (gray dashed line). However, when the radius of antipredator response increases, predation risk decreases. For example, according to the results presented in Fig.~\ref{fig6}: for $R=2$, the reduction in the predation risk is $48.5\%$ (purple dashed line). Our findings also indicate that for $R \geq 3$, the reduction predation risk does not change substantially: $\xi = 56.0\%$, $\xi = 58.1\%$, and $\xi = 59.2\%$ (red dashed line), for $R=3$, $R=4$, and $R=5$, respectively. This means that the benefits of the mobility-limiting antipredator response do not grow significantly for $R>5$; for this reason, we have concentrated our investigation in simulating the cases for $1\,\leq\,R\,\leq\,5$.

Generally speaking, we conclude that the less localized the antipredator opposition is, the more probable the prey escaping predation. This happens because for larger $R$, the average size of the single-species spatial domains increases, as shown in Fig. \ref{fig3}. Because a predation interaction is implemented when a predator has prey as one of its eight immediate neighbors, a less localized antipredator response provides better topological protection to most prey that stays away from the borders of the single-species domains.

\begin{figure}[t]
	\centering
	\includegraphics*[width=8.5cm]{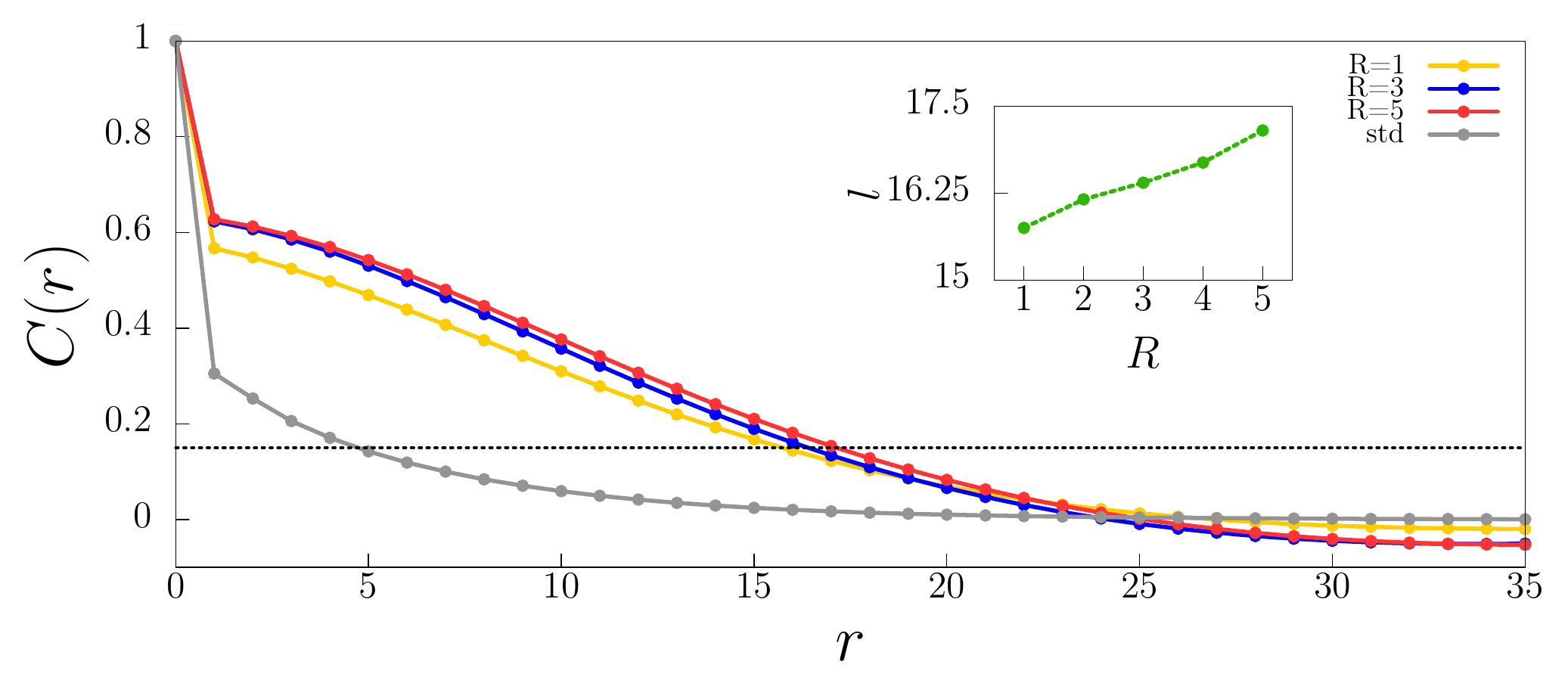}
\caption{Autocorrelation functions $C(r)$. The gray, yellow, blue, and red lines depict the results for the standard model, $R=1$, $R=3$, and $R=5$, respectively. The horizontal dashed black line shows the threshold assumed to calculate the characteristic length. The inset shows the characteristic length in terms of $R$.}
 \label{fig5}
\end{figure}
\section{Autocorrelation Function}

Now we investigate the scale of spatial aggregation of organisms of the same species. For this, we compute the spatial autocorrelation function. Again, assuming the symmetry among the species, we focus only on the spatial segregation of species $1$.

The autocorrelation function is computed from the inverse Fourier transform of
the spectral density as
\be
C(\vec{r}') = \frac{\mathcal{F}^{-1}\{S(\vec{k})\}}{C(0)},
\ee
where $S(\vec{k})$ is given by
\be
S(\vec{k}) = \sum_{k_x, k_y}\,\varphi(\vec{\kappa}),
\ee
and $\varphi(\vec{\kappa})$ is Fourier transform
\be
\varphi(\vec{\kappa}) = \mathcal{F}\,\{\phi(\vec{r})-\langle\phi\rangle\}.
\ee 
The function $\phi(\vec{r})$ represents the spatial distribution of individuals of species $1$ ($\phi(\vec{r})=0$ and $\phi(\vec{r})=1$ indicate the absence and the presence of an individual of species $1$ in at the position $ \vec{r}$ in the lattice, respectively). The spatial autocorrelation function is given by
\be
C(r') = \sum_{|\vec{r}'|=x+y} \frac{C(\vec{r}')}{min [2N-(x+y+1), (x+y+1)]}.
\ee
Moreover, we compute the spatial domains' scale for $C(l)=0.15$, where $l$ is the characteristic length.

Figure 7 depicts the spatial autocorrelation function as a function of the radial coordinate $r$, for the standard model (gray line), $R=1$ (yellow line), $R=3$ (blue line), and $R=5$ (red line). The results were averaged from a set of $100$ simulations with different random initial conditions, running in lattices with $\mathcal{N}=300^2$. The spatial configuration was captured after $3000$ generations, for $p=m=1/2$.

The horizontal black line represents the threshold considered to calculate the length scale, $C(l)\, =\, 0.15$. The outcomes reveal that the spatial clustering of individuals of the same species grows with $R$, as depicted in the inset figure - for the standard case, $l=4.74$. This result confirms that the less localized the antipredator is, the larger the spatial domain's average size. 

\begin{figure}[t]
	\centering
	\includegraphics*[width=8.5cm]{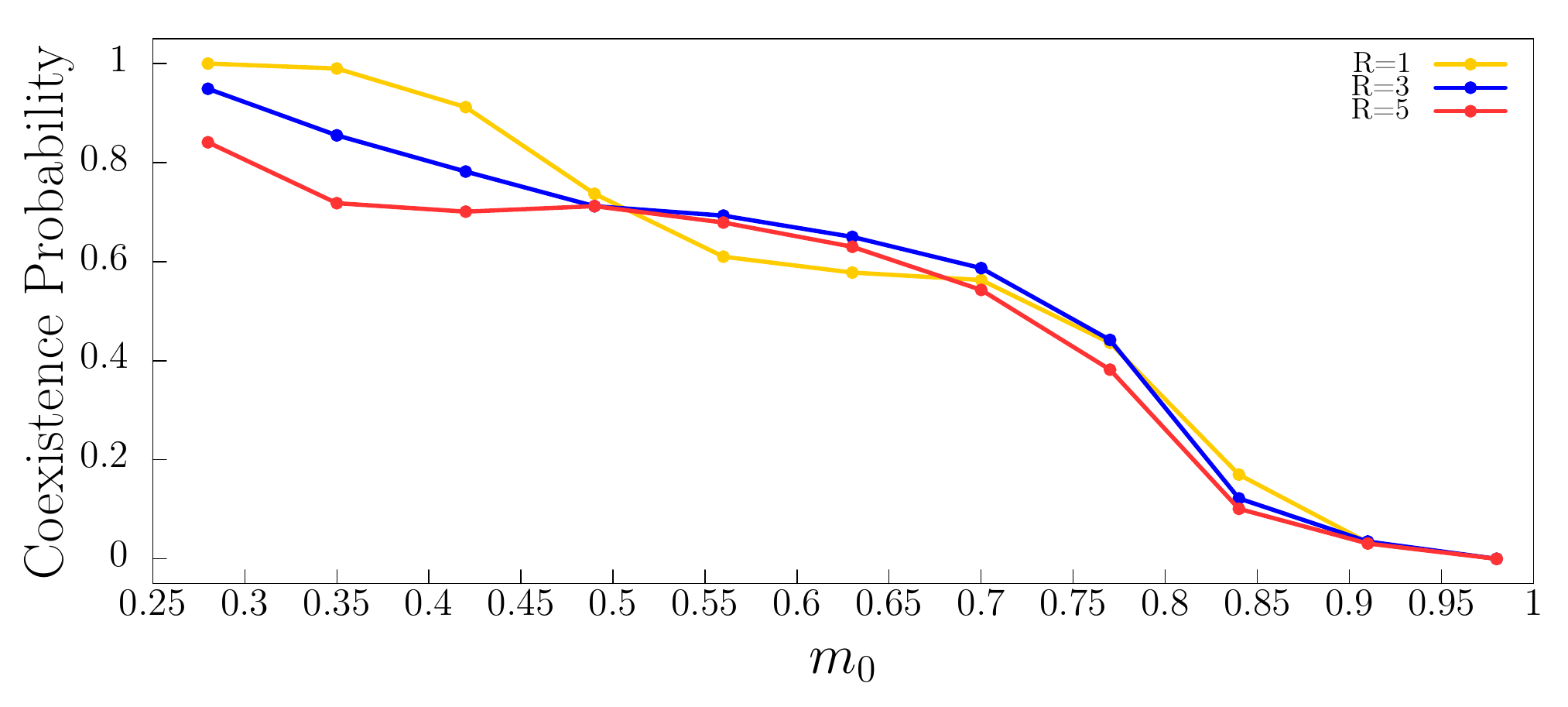}
\caption{Coexistence probability as a function of the mobility probability $m$. The yellow, blue, red, and green lines show the results for $R=1$, $R=3$, $R=5$, respectively.
The results were obtained by running $1000$ simulations in lattices with $120^2$ grid points running until  $120^2$ generations.}
 \label{fig7}
\end{figure}

\section{Coexistence Probability}

Finally, we aim to discover how the locality of the antipredator response affects species diversity. To this purpose, we performed $1000$ simulations in lattices with $120^2$ grid points for $ 0.05\,<\,m_0\,<\,0.95$ in intervals of $ \Delta\, m_0\, =\,0.05$; predation probability was set to $p_0\,=\,1-m_0$. The simulations started from different random initial conditions and ran until $120^2$ generations. Coexistence occurs if all species are present at the end of the simulation. In other words, at least one individual of each species must be present: $I_i (t=120^2) \neq 0$ with $i=1,2,3$.   Otherwise, the simulation results in extinction. We define the coexistence probability as the fraction of realizations resulting in coexistence.

The outcomes are presented in Fig.~$8$, where the
yellow, blue, and red lines show the coexistence probability for $R=1$, $R=3$, and $R=5$, respectively. Overall, species biodiversity is more threatened for higher mobility probabilities. However, the results revealed that for $m_0 \leq 0.5$, coexistence is less probable for a long-range antipredator response. This happens because a short-range antipredator reaction demands a higher individual cost, imposing more substantial limits on the organism's mobility probability.

\section{Comments and Conclusions}

We studied the impact of a mobility-limiting antipredator response on nonhierarchical tritrophic predator-prey systems. Performing stochastic simulations of the rock-paper-scissors model, we assumed that predation probability decreases exponentially with the prey group size. The individual effort devoted to vigilance and collective antipredator tactics decreases an organism's effective mobility probability.
Considering the cost depends on the radius of antipredator response, we investigated the effects on the spatial patterns, population dynamics, and species persistence.

Our results unveiled that the characteristic length of the spatial domains increases for a less localized antipredator response, where individuals can affect the predator's action from longer distances.
This means that the average size of the single-species areas increases for a long-range antipredator response.	
Calculating the predation risk, we found that if the antipredator response is less localized, the prey's vulnerability decreases because more prey may stay away from predators. 

Finally, we studied the influence of the mobility-limiting antipredator response on species diversity. Overall, biodiversity is jeopardized for higher mobility probabilities, irrespective of how localized is the antipredator response is. Moreover, we found that that a more localized antipredator reaction, which demands a higher individual cost in terms of limiting mobility, benefits the maintenance of biodiversity. However, our findings show that the positive effects on the coexistence probability due to the constraining of individuals' fitness do not hold for high mobility probability.

\section{Acknowledgments}
We thank CNPq, ECT, Fapern, and IBED for financial and technical support.

\bibliography{ref}

\end{document}